# SU(5) GRAND UNIFIED THEORY, ITS POLYTOPES AND 5-FOLD SYMMETRIC APERIODIC TILING

MEHMET KOCA[a]

[a] *Retired professor*
mehmetkocaphysics@gmail.com

NAZIFE OZDES KOCA[b*] and ABEER AL-SIYABI[b,**]

[b] *Department of Physics, College of Science, Sultan Qaboos University, P.O. Box 36, Al-Khoud, 123 Muscat, Sultanate of Oman*
* nazife@squ.edu.om
**m21168@student.squ.edu.om

We associate the lepton-quark families with the vertices of the 4D polytopes 5-cell $(0001)_{A4}$ and the rectified 5-cell $(0100)_{A4}$ derived from the $SU(5)$ Coxeter-Dynkin diagram. The off-diagonal gauge bosons are associated with the root polytope $(1001)_{A4}$ whose facets are tetrahedra and the triangular prisms. The edge-vertex relations are interpreted as the $SU(5)$ charge conservation. The Dynkin diagram symmetry of the $SU(5)$ diagram can be interpreted as a kind of particle-antiparticle symmetry. The Voronoi cell of the root lattice consists of the union of the polytopes $(1000)_{A4} + (0100)_{A4} + (0010)_{A4} + (0001)_{A4}$ whose facets are 20 rhombohedra. We construct the Delone (Delaunay) cells of the root lattice as the alternating 5-cell and the rectified 5-cell, a kind of dual to the Voronoi cell. The vertices of the Delone cells closest to the origin consist of the root vectors representing the gauge bosons. The faces of the rhombohedra project onto the Coxeter plane as thick and thin rhombs leading to Penrose-like tiling of the plane which can be used for the description of the 5-fold symmetric quasicrystallography. The model can be extended to $SO(10)$ and even to $SO(11)$ by noting the Coxeter-Dynkin diagram embedding $A_4 \subset D_5 \subset B_5$. Another embedding can be made through the relation $A_4 \subset D_5 \subset E_6$ for more popular $GUT's$. Appendix A includes the quaternionic representations of the Coxeter-Weyl groups $W(A_4) \subset W(H_4)$ which can be obtained directly from $W(E_8)$ by projection. This leads to relations of the $SU(5)$ polytopes with the quasicrystallography in 4D and $E_8$ polytopes. Appendix B discusses the branching of the polytopes in terms of the irreducible representations of the Coxeter-Weyl group $W(A_4) \approx S_5$.

*Keywords*: $SU(5)$ GUT, Polytope, Voronoi cell, Delone cells, Coxeter groups

## 1. Introduction

The Lie algebras and thereof Lie groups derived from the root systems of the Coxeter-Weyl groups are well known: predictions of the standard model of the High Energy Physics described by the Lie group $SU(3) \times SU(2) \times U(1)$ [1,2,3] are heavily based on the Coxeter-Weyl group $W(A_2) \times W(A_1)$. The skeletons [4] of the Grand Unified Theories (GUT), $SU(5) \approx E_4$ [5], $SO(10) \approx E_5$ [6] and the exceptional group $E_6$ [7] are the respective Coxeter-Weyl groups $W(A_4), W(D_5)$ and $W(E_6)$. The charges of the quarks and leptons as well as the gauge bosons obtained from the embedding $W(A_2) \times W(A_1) \subset W(A_4) \approx S_5$ depend on the root and weight vectors which generate the associated lattices invariant under the affine Coxeter group $W_a(A_4)$. The Coxeter number $h$ seems to be relevant to the aperiodic tiling of the Coxeter plane by projection of its root and weight lattices leading to $h$-fold symmetry, a phenomenon which might be relevant to quasicrystallography [8, 9]. The same mathematical technique, namely, the Coxeter-Dynkin diagrams and Coxeter-Weyl groups, can be applied to two totally irrelevant phenomena: the grand unification of the strong and the electroweak interactions and the quasicrystallography, the latter is a mathematically intriguing topic in condensed matter physics. Let us recall that the spontaneous symmetry breaking (Higgs mechanism) and the phase transitions such as super conductivity and super fluidity in condensed matter physics are based on similar mathematical techniques, namely, breaking a group into its subgroups. We point out that the weight vectors of the irreducible representations $\underline{5}^* + \underline{10}$ of $SU(5)$ represent the polytopes 5-cell and the rectified 5-cell respectively. The 5-cell consists of 5 tetrahedra as facets with 5 vertices. The rectified 5-cell consists of 5 tetrahedra and 5 octahedra with 10 vertices. The set of root vectors corresponding to gauge bosons determine the Delone (Delaunay) cells of the root lattice. The union of the irreducible representations $\underline{5}^* + \underline{10} + \underline{10}^* + \underline{5}$ form 30 vertices of the Voronoi cell of the root lattice which is dual to the root polytope with 20 vertices consisting of the cells of 10 tetrahedra and 20 triangular prisms [10]. The Dynkin- Diagram symmetry which extends the Coxeter-Weyl group to the automorphism group of the root system can be related to the charge-conjugation operator.

The paper is organized as follows. In Section 2 we introduce the concept of Voronoi cell and Delone cell by giving the example of the lattice of the affine group $W_a(A_2)$, the point group $W(A_2) \approx S_3$ of which, is the symmetry of the $SU(3)$ polytopes (hexagons and triangles). It will then be straightforward to discuss the $SU(5)$ polytopes and their dual polytopes. Section 3 is devoted to the projection of the $SU(5)$ polytopes into the Coxeter plane. We discuss identification of the charges of the lepton-quark families and the gauge bosons with vertex vectors and the edge vectors respectively. The 5-cell and the rectified 5-cell allow a diagrammatic representation of the interaction of the particles with the gauge bosons in the root and weight spaces analogous to the Feynman diagrams. This is perhaps another aspect of polytopes in particle physics similar to the amplituhedron [11] relating the scattering amplitude to the volume of a certain polytope. Section 4 deals with the projection of the Voronoi cell onto the Coxeter plane leading to Penrose-like tiling of the $SU(5)$ lattice. In the concluding Section 5 we discuss the extension of the work to the GUT models such as $SO(10) \subset SO(11)$, the Coxeter-Weyl group of which is $W(D_5) \subset W(B_5)$ and $E_6$. In appendix A, we introduce the quaternionic representation of the group $W(A_4)$ useful for the embedding $W(A_4) \subset W(H_4)\ ) \subset W(E_8)$ [12, 13]. Appendix B deals with the branching of the $W(A_4)$ orbits of interest in terms of its irreducible representations of the symmetric group $S_5$.

## 2. Polytopes of SU(5), Voronoi and Delone Cells Of The Root Lattice

One should distinguish the unitary symmetry $SU(5)$ which describes an irreducible representation in the Hilbert space describing the wave functions of the particles and the Coxeter-Weyl group $W(A_4) \approx S_5$ which is the discrete symmetry of the root system of $SU(5)$. To set the scene we discuss the polytopes (triangles and hexagons) of $SU(3)$ as an example. The Cartan matrix and its inverse of any Coxeter-Weyl group is defined by the simple roots and weights as

$$C_{ij} = \frac{2(\alpha_i,\alpha_j)}{(\alpha_j,\alpha_j)}, \quad \left(\frac{2\alpha_i}{(\alpha_i,\alpha_i)}, \omega_j\right) = \delta_{ij}, \quad G_{ij} = (C^{-1})_{ij}\frac{(\alpha_j,\alpha_j)}{2} = (\omega_i,\omega_j) \tag{1}$$

where $\alpha_i$ and $\omega_i$ are the simple root vector of the Lie algebra and the corresponding weight vector respectively. The Coxeter group generated by reflections is simply defined as an abstract group [14,15,16,17]

$$< r_1, r_2, \dots, r_n | (r_i r_j)^{m_{ij}} = 1 > . \tag{2}$$

Here $r_i$ is the reflection generator with respect to the hyperplane orthogonal to the root vector $\alpha_i$ which acts on an arbitrary vector as

$$r_i\lambda = \lambda - \frac{2(\alpha_i,\lambda)\alpha_i}{(\alpha_i,\alpha_i)} . \tag{3}$$

Denote an arbitrary vector in the weight space by $\lambda = (a_1\omega_1 + a_2\omega_2 + \dots + a_n\omega_n)$ [18] where $a_i \in \mathbb{Z}$, $i = 1,2,\dots,n$. In this section, we shall consider the Coxeter-Dynkin diagrams of $A_n -$series with the Coxeter-Weyl group $W(A_n) \approx S_{n+1}$. The orbit of the Coxeter-Weyl group will be denoted by $W(A_n)\lambda :=$ $(a_1 a_2 \dots a_n)_{A_n}$ which represents a set of vertices of the polytope of the $W(A_n)$, or shortly, the polytope of $SU(n+1)$.

The $SU(3)$ polytopes $(10)_{A_2}$ and $(01)_{A_2}$ represent two dual triangles and the polytope $(11)_{A_2}$ is a regular hexagon. The root lattice of $SU(3)$ consists of the vectors $(a_1\alpha_1 + a_2\alpha_2)$ with $a_i \in \mathbb{Z}, i = 1,2$ where the unit cell is represented by the hexagon $(11)_{A_2}$. The Voronoi cell of the root lattice is the scaled dual hexagon represented by the union of the triangles $(10)_{A_2} + (01)_{A_2}$. Delaunay (Delone) cells are defined by the root vectors including the origin, a kind of dual to the Voronoi cell [19]. The Delone cells are triangles whose centers are the vertices of the Voronoi cell as shown in Fig. 1.

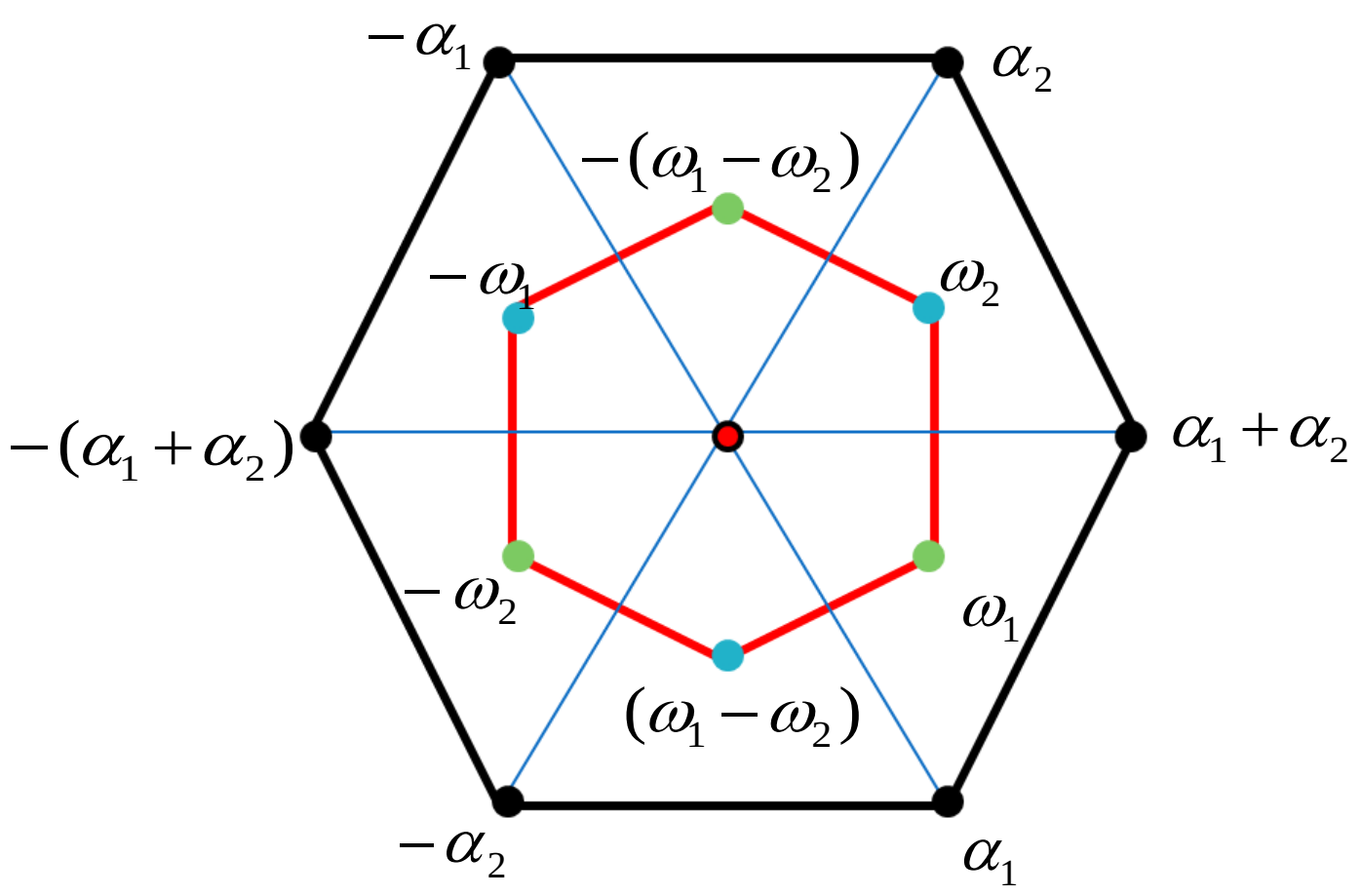

Fig. 1. Voronoi cell (in red) and Delone cells of $SU(3)$ (triangles) whose centers are the vertices of the Voronoi cell.

Since $SU(3)$ will be taken as the color subgroup of $SU(5)$ we label its 3-dimensional irreducible representation as *r* (red), *b* (blue) and *g* (green). The color charges of the quarks or the gluons can be computed by taking the scalar products of the weight and/or root vectors with the color hypercharge and color charge vectors represented respectively by $Y^c = \omega_2$ and $Q^c = \omega_1$ where the color isospin vector is given by $I_3^c = \omega_1 - \frac{1}{2}\omega_2$. For example, the color isospin and color hypercharge $(I_3^c, Y^c)$ of the quark represented by the weight $\omega_1$ is given by $\left(\frac{1}{2}, \frac{1}{3}\right)$, a state represented by *r* (red).

Similarly, the orbits of $SU(4)$ $(100)_{A_3}$, $(001)_{A_3}$ represent two dual tetrahedra while $(010)_{A_3}$, $(101)_{A_3}$ and $(111)_{A_3}$, represent octahedron, cubeoctahedron and truncated octahedron respectively [20]. The root lattice of $(A_3)$, also called fcc lattice, has the Voronoi cell as the rhombic dodecahedron represented by the union of the orbits $(100)_{A_3}$+$(010)_{A_3}$ + $(001)_{A_3}$. The Delone cells of the root lattice consist of the alternating tetrahedra and octahedra taking the vertices from the root lattice including the origin. The Voronoi cell of the weight lattice ( reciprocal lattice of the root lattice) is the truncated octahedron $\frac{1}{4}(111)_{A_3}$.

After these preliminary examples, we can discuss the polytopes of $SU(5)$ corresponding to the quark-lepton families and the gauge bosons of $SU(5)$. The Coxeter-Dynkin diagram of $SU(5)$ is shown in Fig. 2.

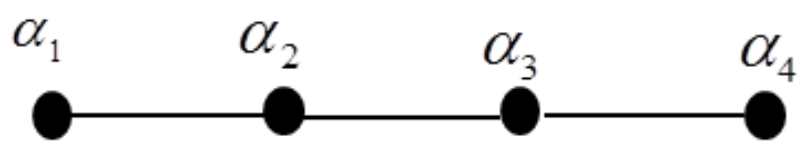

Fig. 2. Coxeter-Dynkin diagram of $SU(5)$.

The Cartan matrix and its inverse are given as

$$C = \begin{pmatrix} 2 & -1 & 0 & 0 \\ -1 & 2 & -1 & 0 \\ 0 & -1 & 2 & -1 \\ 0 & 0 & -1 & 2 \end{pmatrix}, \qquad C^{-1} = \frac{1}{5}\begin{pmatrix} 4 & 3 & 2 & 1 \\ 3 & 6 & 4 & 2 \\ 2 & 4 & 6 & 3 \\ 1 & 2 & 3 & 4 \end{pmatrix}. \tag{4}$$

The $\sqrt{detC} = \sqrt{5}$ represents the volume of the Voronoi cell or the parallelotope generated by the simple roots. It is sometimes more convenient to work with the orthonormal set of vectors to represent the simple roots $\alpha_1 = l_1 - l_2, \alpha_2 = l_2 - l_3, \alpha_3 = l_3 - l_4, \alpha_4 = l_4 - l_5$ where $(l_i, l_j) = \delta_{ij}$. The generators $r_i, i = 1,2,3,4$ of the Coxeter-Weyl group $W(A_4) \approx S_5$ of order 120 act on the weight vectors as $r_i\omega_j = \omega_j - \delta_{ij}\alpha_j$. The orbit $(1000)_{A_4} = \{\omega_1, \omega_2 - \omega_1, \omega_3 - \omega_2, \omega_4 - \omega_3, -\omega_4\}$ [10] constitutes the weights of the irreducible representation $\underline{5}$ of $SU(5)$. The weight vectors represent the vertices of the polytope 5-cell which consists of five tetrahedral facets, four of which meeting at one vertex. The symmetric group $S_5$ permutes the vertices of 5-cell as well as its facets. The numbers of vertices $N_0 = 5$, edges $N_1 = 10$, faces $N_2 = 10$ and the cells $N_3 = 5$ satisfy the Euler equation $N_0 - N_1 + N_2 - N_3 = 0$. The orbit $(0100)_{A_4}$ constitutes weight vectors of the irreducible representation $\underline{10}$ of $SU(5)$, so called rectified 5-cell, with the respective 10 vertices, 30 edges, 30 faces (triangles) and 10 cells (5 tetrahedra+5octahedra). The weight vectors of the orbit $(0010)_{A_4}$ which represent the irreducible representation $\underline{10^*}$ and the orbit $(0001)_{A_4} = \underline{5^*}$ can be respectively obtained from the orbits $(0100)_{A_4}$ and $(1000)_{A_4}$by applying the Dynkin-diagram symmetry which exchanges the weights $\gamma: \omega_1 \leftrightarrow \omega_4,\ \omega_2 \leftrightarrow \omega_3$. With the Dynkin diagram symmetry the Coxeter-Weyl group of $SU(5)$ can be extended to the group $Aut(A_4) \approx W(A_4){:}C_2 \approx S_5{:}S_2$ of order 240. The $SU(5)$ charges of the quark-lepton families can be obtained from the weight vectors of the orbits $(0001)_{A_4} = \underline{5^*}$ and $(0100)_{A_4} = \underline{10}$. The orbit $(1001)_{A_4}$ (runcinated 5-cell; or, better, call it the root polytope) consists of 20 vertices, 60 edges, 70 faces (40 triangles+30 squares) and 30 facets (10 tetrahedra+20 triangular prisms). More information about the $SU(5)$ polytopes can be found in the reference [10]. Vertices of the root polytope are given as $\pm(l_i - l_j), i \neq j$ or more explicitly,

$$\pm(\alpha_1 + \alpha_2 + \alpha_3 + \alpha_4),\ \pm(\alpha_1 + \alpha_2 + \alpha_3),\ \pm(\alpha_2 + \alpha_3 + \alpha_4),$$

$$\pm(\alpha_1 + \alpha_2),\ \pm(\alpha_2 + \alpha_3),\ \pm(\alpha_3 + \alpha_4),\ \pm\alpha_1,\ \pm\alpha_2,\ \pm\alpha_3,\ \pm\alpha_4. \tag{5}$$

Moreover, we have quadruply degenerate zero root vectors. In the Lie algebraic terminology, zero roots correspond to the Cartan generators of the Lie algebra and in the adjoint representation we have four gauge bosons all with zero $SU(5)$ charges.

One can define the color isospin vector $I_3^c$, color hypercharge $Y^c$, weak isospin $I_3^W$and weak hypercharge $Y^W$ in terms of the weight and root vectors as

$$I_3^c = \frac{1}{2}(\omega_1 + \omega_2 - \omega_3) = \frac{1}{2}(\alpha_1 + \alpha_2) = \frac{1}{2}(l_1 - l_3),$$

$$Y^c = \frac{1}{3}(\omega_1 - \omega_2 + \omega_3 + 2\omega_4) = \frac{1}{3}(\alpha_1 + \alpha_2 + 2\alpha_3 + 2\alpha_4) = \frac{1}{3}(l_1 + l_3 - 2l_5),$$

$$I_3^W = \frac{1}{2}(-\omega_1 + \omega_2 + \omega_3 - \omega_4) = \frac{1}{2}(\alpha_2 + \alpha_3) = \frac{1}{2}(l_2 - l_4), \tag{6}$$

$$Y^W = \frac{5}{3}(-\omega_1 + \omega_2 - \omega_3 + \omega_4) = \frac{1}{3}(-2\alpha_1 + \alpha_2 - \alpha_3 + 2\alpha_4) = \frac{1}{3}(-2l_1 + 3l_2 - 2l_3 + 3l_4 - 2l_5).$$

Note that these 4 vectors are orthogonal to each other. Let us give a few examples for the determination of the particles corresponding to the weights or roots in the representations $\underline{5^*}$ , $\underline{10}$ and the orbit $(1001)_{A_4}$ . The quantum numbers of the particles corresponding to the weights and roots $\omega_4$, $\omega_2$, $\pm\alpha_1$ and $\pm(\alpha_1 +$ $\alpha_2+\alpha_3+\alpha_4$) in the order of equation (6) are $(0,\frac{2}{3},0,\frac{2}{3})$: $(\overline{d_g})_L$; $(\frac{1}{2},\frac{1}{3},\frac{1}{2},\frac{1}{3})$: $(u_r)_L$; $\pm(\frac{1}{2},\frac{1}{3},-\frac{1}{2},-\frac{5}{3})$: $(X_r,\overline{X_r})$ and $\pm(\frac{1}{2},1,0,0)$: $(G_{r\bar{g}},\overline{G_{r\bar{g}}})$ respectively. Here $X_r$ and $G_{r\bar{g}}$ represent the $SU(5)$ gauge boson and the gluon respectively. All the particles and gauge bosons of $SU(5)$ and the corresponding weights and roots are listed in Table 1, Table 2 and Table 3.

Table 1. Weight vectors of $\underline{5^*}$ and the corresponding particles.

| Vectors | $I_3^c$ | $Y^c$ | $I_3^W$ | $Y^W$ | $Q^{em}$ | particle |
|---|---|---|---|---|---|---|
| $\omega_3-\omega_4$ | 0 | 0 | $\frac{1}{2}$ | $-1$ | 0 | $(\nu_e)_L$ |
| $\omega_1-\omega_2$ | 0 | 0 | $-\frac{1}{2}$ | $-1$ | $-1$ | $(e)_L$ |
| $\omega_4$ | 0 | $\frac{2}{3}$ | 0 | $\frac{2}{3}$ | $\frac{1}{3}$ | $(\overline{d_g})_L$ |
| $\omega_2-\omega_3$ | $\frac{1}{2}$ | $-\frac{1}{3}$ | 0 | $\frac{2}{3}$ | $\frac{1}{3}$ | $(\overline{d_b})_L$ |
| $-\omega_1$ | $-\frac{1}{2}$ | $-\frac{1}{3}$ | 0 | $\frac{2}{3}$ | $\frac{1}{3}$ | $(\overline{d_r})_L$ |

Table 2. Weight vectors of $\underline{10}$ and the corresponding particles.

| Vectors | $I_3^c$ | $Y^c$ | $I_3^W$ | $Y^W$ | $Q^{em}$ | particle |
|---|---|---|---|---|---|---|
| $\omega_2$ | $\frac{1}{2}$ | $\frac{1}{3}$ | $\frac{1}{2}$ | $\frac{1}{3}$ | $\frac{2}{3}$ | $(u_r)_L$ |
| $-\omega_1+\omega_3$ | $-\frac{1}{2}$ | $\frac{1}{3}$ | $\frac{1}{2}$ | $\frac{1}{3}$ | $\frac{2}{3}$ | $(u_b)_L$ |
| $-\omega_1+\omega_2-\omega_4$ | 0 | $-\frac{2}{3}$ | $\frac{1}{2}$ | $\frac{1}{3}$ | $\frac{2}{3}$ | $(u_g)_L$ |
| $\omega_1-\omega_2+\omega_3$ | 0 | $\frac{2}{3}$ | 0 | $-\frac{4}{3}$ | $-\frac{2}{3}$ | $(\overline{u_g})_L$ |
| $\omega_1-\omega_4$ | $\frac{1}{2}$ | $-\frac{1}{3}$ | 0 | $-\frac{4}{3}$ | $-\frac{2}{3}$ | $(\overline{u_b})_L$ |
| $-\omega_2+\omega_3-\omega_4$ | $-\frac{1}{2}$ | $-\frac{1}{3}$ | 0 | $-\frac{4}{3}$ | $-\frac{2}{3}$ | $(\overline{u_r})_L$ |
| $\omega_1-\omega_3+\omega_4$ | $\frac{1}{2}$ | $\frac{1}{3}$ | $-\frac{1}{2}$ | $\frac{1}{3}$ | $-\frac{1}{3}$ | $(d_r)_L$ |
| $-\omega_2+\omega_4$ | $-\frac{1}{2}$ | $\frac{1}{3}$ | $-\frac{1}{2}$ | $\frac{1}{3}$ | $-\frac{1}{3}$ | $(d_b)_L$ |
| $-\omega_3$ | 0 | $-\frac{2}{3}$ | $-\frac{1}{2}$ | $\frac{1}{3}$ | $-\frac{1}{3}$ | $(d_g)_L$ |
| $-\omega_1+\omega_2-\omega_3+\omega_4$ | 0 | 0 | 0 | 2 | 1 | $(e^+)_L$ |

It is interesting to note that the edges of the polytopes correspond to the gauge bosons. Let us consider the edges of the polytope $\underline{5^*}$ connecting the vertex $\omega_4$ to the rest of the vertices. They can be represented, depending on the direction of the connecting vector, as root vectors:
$\pm\alpha_4,\ \pm(\alpha_4+\alpha_3),\ \pm(\alpha_4+\alpha_3+\alpha_2),\ \pm(\alpha_4+\alpha_3+\alpha_2+\alpha_1)$ .

This also indicates that 4 edges intersect at the same vertex.

Table 3. Root system of $SU(5)$ and the corresponding gauge bosons.

| Root Vectors | $Y^c$ | $I_3^c$ | $Y^W$ | $I_3^W$ | $Q^{em}$ | Particle |
|---|---|---|---|---|---|---|
| $(\alpha_1+\alpha_2+\alpha_3+\alpha_4)$ | 1 | $1/2$ | 0 | 0 | 0 | $G_{r\bar{g}}$ |
| $-(\alpha_1+\alpha_2+\alpha_3+\alpha_4)$ | $-1$ | $-1/2$ | 0 | 0 | 0 | $G_{g\bar{r}}$ |
| $(\alpha_2+\alpha_3+\alpha_4)$ | $2/3$ | 0 | $5/3$ | $1/2$ | $4/3$ | $\overline{X_g}$ |
| $-(\alpha_2+\alpha_3+\alpha_4)$ | $-2/3$ | 0 | $-5/3$ | $-1/2$ | $-4/3$ | $X_g$ |
| $(\alpha_3+\alpha_4)$ | 1 | $-1/2$ | 0 | 0 | 0 | $G_{b\bar{g}}$ |
| $-(\alpha_3+\alpha_4)$ | $-1$ | $1/2$ | 0 | 0 | 0 | $G_{\bar{b}g}$ |
| $(\alpha_4)$ | $2/3$ | 0 | $5/3$ | $-1/2$ | $1/3$ | $\overline{Y_g}$ |
| $-(\alpha_4)$ | $-2/3$ | 0 | $-5/3$ | $1/2$ | $-1/3$ | $Y_g$ |
| $(\alpha_3)$ | $1/3$ | $-1/2$ | $-5/3$ | $1/2$ | $-1/3$ | $Y_b$ |
| $-(\alpha_3)$ | $1/3$ | $1/2$ | $5/3$ | $-1/2$ | $1/3$ | $\bar{Y}_b$ |
| $(\alpha_2+\alpha_3)$ | 0 | 0 | 0 | 1 | 1 | $W^+$ |
| $-(\alpha_2+\alpha_3)$ | 0 | 0 | 0 | $-1$ | $-1$ | $W^-$ |
| $(\alpha_1+\alpha_2+\alpha_3)$ | $1/3$ | $1/2$ | $-5/3$ | $1/2$ | $-1/3$ | $Y_r$ |
| $-(\alpha_1+\alpha_2+\alpha_3)$ | $-1/3$ | $-1/2$ | $5/3$ | $-1/2$ | $1/3$ | $\bar{Y}_r$ |
| $(\alpha_2)$ | $-1/3$ | $1/2$ | $5/3$ | $1/2$ | $4/3$ | $\overline{X_b}$ |
| $-(\alpha_2)$ | $1/3$ | $-1/2$ | $-5/3$ | $-1/2$ | $-4/3$ | $X_b$ |
| $(\alpha_1+\alpha_2)$ | 0 | 1 | 0 | 0 | 0 | $G_{r\bar{b}}$ |
| $-(\alpha_1+\alpha_2)$ | 0 | $-1$ | 0 | 0 | 0 | $G_{\bar{r}b}$ |
| $(\alpha_1)$ | $1/3$ | $1/2$ | $-5/3$ | $-1/2$ | $-4/3$ | $X_r$ |
| $-(\alpha_1)$ | $-1/3$ | $-1/2$ | $5/3$ | $1/2$ | $4/3$ | $\overline{X_r}$ |
| 0 | 0 | 0 | 0 | 0 | 0 | $G_{\frac{r\bar{r}-b\bar{b}}{\sqrt{2}}}$ |
| 0 | 0 | 0 | 0 | 0 | 0 | $G_{\frac{r\bar{r}+b\bar{b}-2g\bar{g}}{\sqrt{6}}}$ |
| 0 | 0 | 0 | 0 | 0 | 0 | $A$ |
| 0 | 0 | 0 | 0 | 0 | 0 | $Z$ |

The simple roots of $SU(5)$ generate a lattice $x_1 l_1 + x_2 l_2 + x_3 l_3 + x_4 l_4 + x_5 l_5$ with $x_1 + x_2 + x_3 + x_4 + x_5 = 0$, $x_i \in \mathbb{Z}$ which consists of alternating 5-cell and rectified 5-cell. The orbit $(1001)_{A_4}$ is a convex polytope as we call the root polytope. The dual of the root polytope is the Voronoi cell $V(0)$ of the root lattice whose 30 vertices are the union of the orbits (for details, see reference [10]),

$$\text{Voronoi cell: } (1000)_{A_4} \cup (0100)_{A_4} \cup (0010)_{A_4} \cup (0001)_{A_4}. \quad (7)$$

It has 30 vertices, 70 edges, 60 faces (rhombs) and 20 facets (rhombohedra). A typical rhombohedron shown in Fig. 3 has 8 vertices

$$\omega_1, \omega_4, \omega_2, \omega_3, \; r_2 r_3 \omega_2, r_3 r_2 \omega_2, r_2 r_3 \omega_3, r_3 r_2 \omega_3 \quad (8)$$

whose center is represented by the vector

$$\frac{1}{2}(\omega_1 + \omega_4) = \frac{1}{2}(\alpha_1 + \alpha_2 + \alpha_3 + \alpha_4) = \frac{1}{2}(l_1 - l_5). \quad (9)$$

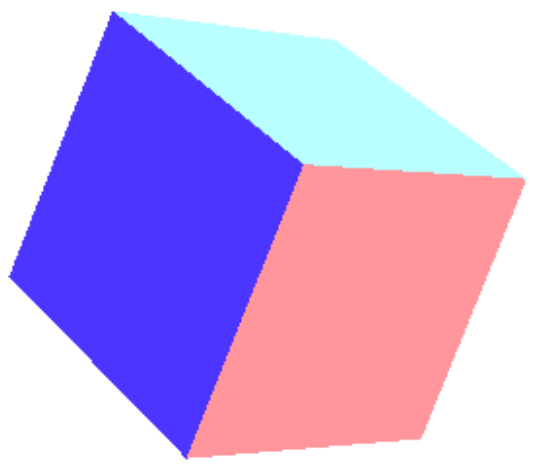

Fig. 3. A rhombohedron as the typical facet of the Voronoi cell.

A typical face of the 6 faces of the rhombohedron is a rhombus which can be represented by 4 vertices $\omega_1, \omega_2, \omega_3, r_2 r_3 \omega_2$ as shown in Fig. 4 with an edge length $\frac{2}{\sqrt{5}}$ and the interior angles about $75.5^\circ$ and $104.5^\circ$.

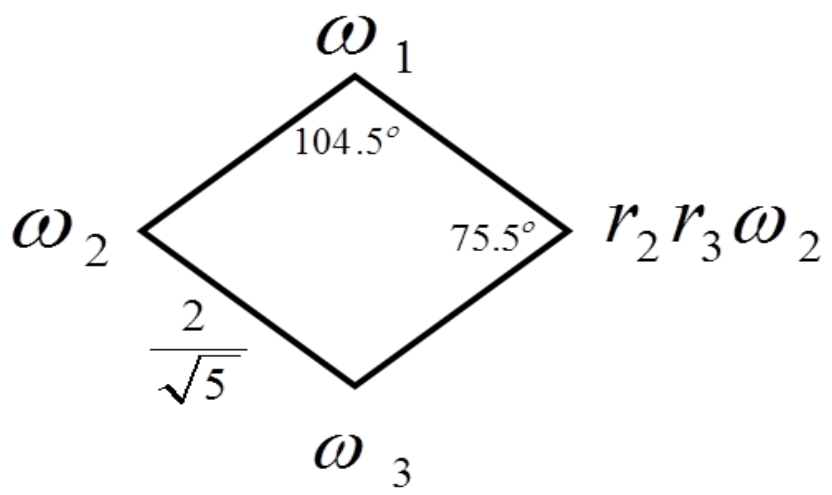

Fig. 4. A rhombic face of the rhombohedron of Fig. 3.

The vector (9) is half the highest weight vector which represents a hyperplane shifted from the origin to the position $\frac{1}{2}(\omega_1 + \omega_4)$; when a reflection generator with respect to this hyperplane is added to the four reflection generators of the group $W(A_4)$ the group is extended to the affine group $W_a(A_4)$ of infinite order. The root and weight lattices are generated by reflections with respect to these 5 mirrors, the fifth is represented by the hyperplane orthogonal to the highest weight vector. There are 20 hyperplanes through the centers of the facets which are the halves of the root vectors listed in (5). Reflecting the Voronoi cell $V(0)$ centered at the origin with respect to the hyperplanes located at the halves of the roots will generate 20 new Voronoi cells surrounding the Voronoi cell $V(0)$. The process of folding at the facets will fill the 4D space of the root lattice with the congruent Voronoi cells. The Voronoi cell of the weight lattice is even more interesting known as the permutohedron with 120 vertices and is represented by the polytope $\frac{1}{5}(1111)_{A_4}$ obtained as the orbit of the scaled Weyl vector. The scale factor 5 here is the Coxeter number of the Coxeter-Weyl group $W(A_4) \approx S_5$.

Next, we will explain how the root lattice is represented as the pattern of alternating 5-cells and rectified 5-cells. To construct the vertices of the 5-cells in the root lattice whose centers are the vertices of the 5-cell determined by the orbit $(1000)_{A_4}$we start with the vertex $\omega_1$. The generators $r_2, r_3, r_4$ leaving the vector $\omega_1$ invariant generate a tetrahedral group of order 24. When the tetrahedral group generated by these generators, for example, is applied to the root $\alpha_1$ it generates four root vectors $\alpha_1, \alpha_1+\alpha_2, \alpha_1+\alpha_2+\alpha_3, \alpha_1+\alpha_2+\alpha_3+\alpha_4$which form a tetrahedron. One can prove that including the origin as a new vertex, five vertices form a 5-cell whose center is $\omega_1$.The 5-cells admitting the other vertices of the polytope $(1000)_{A4}$ and the 5-cells obtained from the orbit $(0001)_{A4}$ are listed in Table 4. The next five Delone cells can be obtained from those in the Table 4 by using the Dynkin diagram symmetry.

Table 4. Delone 5-cells represented by root vectors.

| center of the 5-cell | Vertices of the 5-cell | Applied generators to get the next 5-cell |
|---|---|---|
| $\omega_1$ | $0, \alpha_1, \alpha_1+\alpha_2, \alpha_1+\alpha_2+\alpha_3, \alpha_1+\alpha_2+\alpha_3+\alpha_4$ | $r_1$ |
| $\omega_2-\omega_1$ | $0, -\alpha_1, \alpha_2, \alpha_2+\alpha_3, \alpha_2+\alpha_3+\alpha_4$ | $r_2$ |
| $\omega_3-\omega_2$ | $0, -\alpha_2, \alpha_3, -(\alpha_1+\alpha_2), \alpha_3+\alpha_4$ | $r_3$ |
| $\omega_4-\omega_3$ | $0, -\alpha_3, \alpha_4, -(\alpha_2+\alpha_3), -(\alpha_1+\alpha_2+\alpha_3)$ | $r_4$ |
| $-\omega_4$ | $0, -\alpha_4, -(\alpha_3+\alpha_4), -(\alpha_2+\alpha_3+\alpha_4), -(\alpha_1+\alpha_2+\alpha_3+\alpha_4)$ | |

To give an example for the Delone rectified 5-cell we construct the one whose center is $\omega_2$ which is invariant under the generators $r_1, r_3, r_4$forming a group of symmetry of the prism. The roots $\alpha_2, \alpha_2+\alpha_3, \alpha_2+\alpha_3+\alpha_4$ form a triangle under the generators $r_3, r_4$. If we apply the generator $r_1$on these 3 vertices we obtain the roots $\alpha_1+\alpha_2, \alpha_1+\alpha_2+\alpha_3, \alpha_1+\alpha_2+\alpha_3+\alpha_4$ which form another triangle parallel to the first; together they form a triangular prism. These 6 roots and the lattice vectors $\alpha_1+2\alpha_2+\alpha_3$, $\alpha_1+2\alpha_2+\alpha_3+\alpha_4$, $\alpha_1+2\alpha_2+2\alpha_3+\alpha_4$, including the origin form the Delone rectified 5-cell. As can be checked the average of 10 vertices is $\omega_2$. One can prove that the rectified 5-cell consists of 5 tetrahedra and 5 octahedra as it is expected. Let us enumerate 9 vertices as

$$
\begin{aligned}
&1:\ l_1+l_2-l_4-l_5\\
&2:\ l_1+l_2-l_3-l_5\\
&3:\ l_1+l_2-l_3-l_4\\
&4:\ l_2-l_5\\
&5:\ l_2-l_4\\
&6:\ l_2-l_3\\
&7:\ l_1-l_5\\
&8:\ l_1-l_4\\
&9:\ l_1-l_3
\end{aligned}
\tag{10}
$$

and let 0 denotes the origin.

Then one can prove that each set of four vertices

$$(4,5,6,0), (7,8,9,0), (1,2,4,7), (2,3,6,9) \text{ and } (1,3,5,8)$$

form a tetrahedron. Similarly, each set of 6 vertices

$$(1,4,5,7,8,0), (2,4,6,7,9,0), (3,5,6,8,9,0), (1,2,3,4,5,6) \text{ and } (1,2,3,7,8,9)$$

form an octahedron. Altogether they represent the rectified 5-cell forming a Delone cell in the root lattice surrounding the vector $\omega_2$. The 20 rectified 5-cells can be obtained from the set (10) by using the generators of the group $W(A_4)$ and the Dynkin diagram symmetry.

In the next section, we will discuss the projections of these polytopes onto the Coxeter plane and identify the vertices and edges of the polytopes with the particles and gauge bosons.

## 3. Diagrammatic Representations of The Particles And Gauge Bosons In The Coxeter Plane

Here we first discuss the projection technique of the polytopes onto the Coxeter plane. Consider the Cartan matrix $C$ of a simply laced root system where all roots have the same norm. Let $\mu_i$ be the eigenvalues and $\vec{X}_i$ the corresponding normalized eigenvectors of the Cartan matrix. Earlier we have shown [9] that a set of orthonormal vectors in $n$-dimensions can be obtained as

$$\hat{x}_i = \frac{1}{\sqrt{\mu_i}} \sum_{j=1}^{n} \alpha_j X_{ji},\ i = 1,2,\dots,n. \tag{11}$$

Define the generators $R_1 = r_1 r_3,\ \ R_2 = r_2 r_4$ satisfying the relations $R_1{}^2 = R_2{}^2 = (R_1 R_2)^5 = 1$. They generate the dihedral group of order 10. In the ordered basis $\hat{x}_4, \hat{x}_1, \hat{x}_2, \hat{x}_3$, $R_1 R_2$ is a block-diagonal matrix in the form

$$R = R_1 R_2 = \begin{pmatrix} \cos(\frac{2\pi}{5}) & -\sin(\frac{2\pi}{5}) & 0 & 0 \\ sin(\frac{2\pi}{5}) & \cos(\frac{2\pi}{5}) & 0 & 0 \\ 0 & 0 & \cos(\frac{4\pi}{5}) & -\sin(\frac{4\pi}{5}) \\ 0 & 0 & \sin(\frac{4\pi}{5}) & \cos(\frac{4\pi}{5}) \end{pmatrix}. \tag{12}$$

This shows that the polytopes can be projected into the Coxeter plane defined by the unit vectors $E := (\hat{x}_4, \hat{x}_1)$ with a 5-fold symmetry. In terms of the simple roots and the corresponding weights, the unit vectors can be given as

$$\hat{x}_1 = \frac{\sqrt{2+\tau}}{\sqrt{10}}(\alpha_1 + \tau\alpha_2 + \tau\alpha_3 + \alpha_4) = \frac{1}{\sqrt{2(2+\tau}}(-\sigma\omega_1 + \omega_2 + \omega_3 - \sigma\omega_4),$$

$$\hat{x}_2 = \frac{1}{\sqrt{10}}(-\tau\alpha_1 - \alpha_2 + \alpha_3 + \tau\alpha_4) = \frac{1}{\sqrt{2}}(-\omega_1 + \sigma\omega_2 - \sigma\omega_3 + \omega_4),$$

$$\hat{x}_3 = \frac{\sqrt{2+\tau}}{\sqrt{10}}(\alpha_1 + \sigma\alpha_2 + \sigma\alpha_3 + \alpha_4) = \frac{1}{\sqrt{2(2+\sigma}}(\tau\omega_1 - \omega_2 - \omega_3 + \tau\omega_4),$$

$$\hat{x}_4 = \frac{1}{\sqrt{10}}(-\sigma\alpha_1 - \alpha_2 + \alpha_3 + \sigma\alpha_4) = \frac{1}{\sqrt{2}}(-\omega_1 + \tau\omega_2 - \tau\omega_3 + \omega_4). \tag{13}$$

Here $\tau = \frac{1+\sqrt{5}}{2}$ is the golden ratio and its algebraic conjugate is $\sigma = \frac{1-\sqrt{5}}{2}$.

Let us define the complex parameter $\zeta = e^{i\frac{2\pi}{5}}$ [21]. Then the components of the simple roots and the weights in the Coxeter plane $E = (\hat{x}_4, \hat{x}_1)$ can be written as

$$\alpha_{1E} = \sqrt{\frac{2}{5}}(\zeta - \zeta^2),\ \alpha_{2E} = \sqrt{\frac{2}{5}}(-1 - \zeta^2),\ \alpha_{3E} = \sqrt{\frac{2}{5}}(1 - \zeta^3), \alpha_{4E} = \sqrt{\frac{2}{5}}(\zeta^3 - \zeta^4) \tag{14}$$

$$\omega_{1E}=\sqrt{\frac{2}{5}}\zeta,\ \omega_{2E}=\sqrt{\frac{2}{5}}(\zeta+\zeta^2),\ \omega_{3E}=-\sqrt{\frac{2}{5}}(\zeta^3+\zeta^4), \omega_{4E}=-\sqrt{\frac{2}{5}}\zeta^4. \quad (15)$$

Vertices of the 5-cell of the orbit $(1000)_{A_4}$, arranged as the orbit of the dihedral group $< R_1, R_2 >$, constitutes a pentagon with vertices in the order $\sqrt{\frac{2}{5}}(1,\zeta,\zeta^2,\zeta^3,\zeta^4)$ while the rectified 5-cell $(0100)_{A_4}$ consists of two concentric pentagons first with the coordinates $\sqrt{\frac{2}{5}}[(\zeta+\zeta^2),(\zeta^2+\zeta^3),(\zeta^3+\zeta^4),(\zeta^4+\zeta^5),(\zeta^5+\ \zeta)]$ and the second with the coordinates $\sqrt{\frac{2}{5}}[(1+\zeta^2),\ (\zeta+\zeta^3),(\zeta^2+\zeta^4),(\zeta^3+\zeta^5),(\zeta^4+\ \zeta)]$ with radii $\sqrt{\frac{2}{5}}\,\tau$ and $-\sqrt{\frac{2}{5}}\,\sigma$ respectively. Note that $\zeta^5=1$. The projection of the 5-cell that is the projection of the polytope $\underline{5^*}$ is illustrated in Fig. 5 and that of the rectified 5-cell $\underline{10}$ is given in Fig. 6 [22].

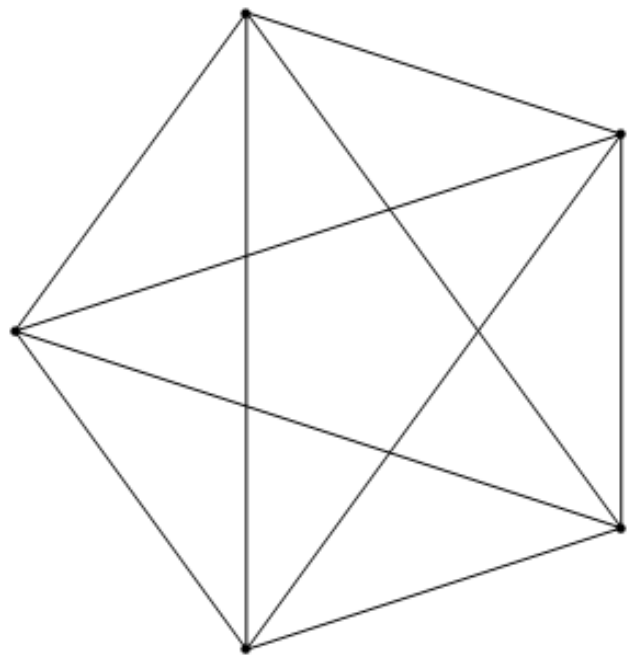

Fig. 5. Projection of the polytope $\underline{5^*}$ .

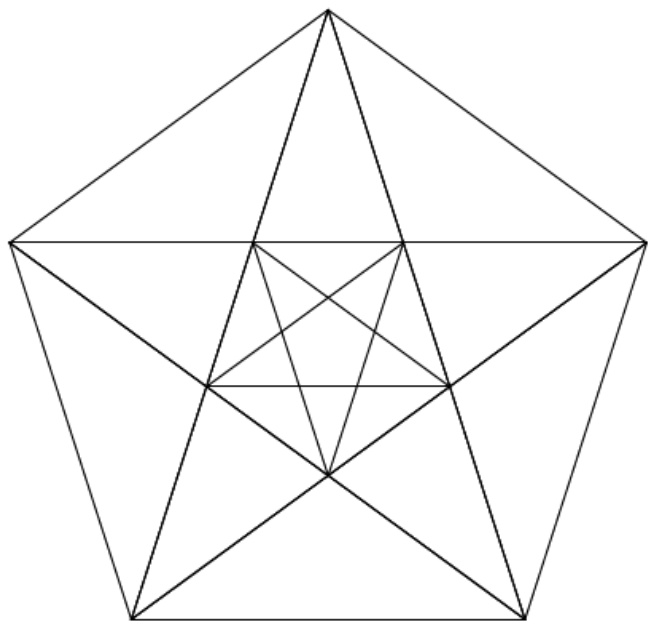

Fig. 6. Projection of the polytope $\underline{10}$.

Now let us discuss the vertex and edge relations in terms of particles and gauge bosons. Take for instance, the weight $\omega_4$ from the orbit $\underline{5^*}$ and apply the operator $R_1R_2$, then one generates the sequence of weights

$$\omega_4, \qquad \omega_2-\omega_3\,, \quad -\omega_1, \qquad \omega_1-\omega_2, \qquad \omega_3-\omega_4. \quad (16a)$$

We list the corresponding quarks and leptons in the same order,

$$(\overline{d_g})_L, \qquad (\overline{d_b})_L, \qquad (\overline{d_r})_L, \qquad e_L^-, \qquad \nu_{eL}. \quad (16b)$$

The four edges joining the vertices, say, to $\nu_{eL}$ in the forward direction are represented by the roots $\alpha_2 + \alpha_3,\ \alpha_1 + \alpha_2 + \alpha_3,\ \alpha_3,\ \alpha_4$ corresponding to the gauge bosons $W^+, Y_r, Y_b, Y_g$ respectively. This implies some similarity to the amplituhedron technique [11] where one obtains a polytope by permuting the momenta of the particles constrained by the conservation of momentum. Here it is the vectors representing edges of the polytope corresponding to the charges of the gauge bosons. The polytope $\underline{5}^*$ whose vertices representing the quarks and leptons are illustrated in Fig. 7.

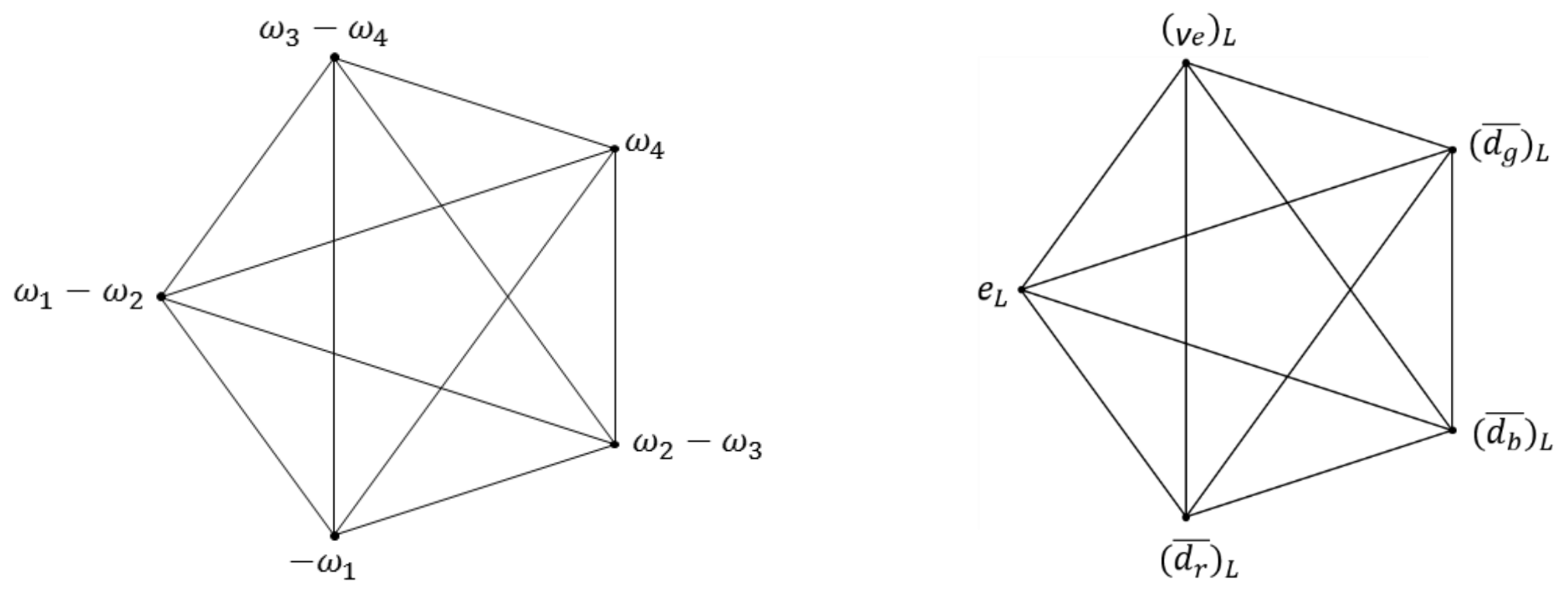

Fig. 7. First generations of the quark-lepton families in the representation $\underline{5}^*$.

To give another example let us take the weight $\omega_2$ from the polytope $\underline{10}$ representing the particle $(u_r)_L$ and determine the gauge bosons which link the particle associated with $\omega_2$ with the other particles in the same orbit. The six weights and six edges connecting $\omega_2$ to the other particles are given by

$$\begin{array}{lll} -\omega_1 + \omega_2 - \omega_3 + \omega_4, & \omega_1 - \omega_2 + \omega_3, & -\omega_1 + \omega_3, \\ \omega_1 - \omega_3 + \omega_4, & -\omega_1 + \omega_2 - \omega_4, & \omega_1 - \omega_4. \end{array} \tag{17}$$

The particles in the same order are

$$\begin{array}{lll} (e^+)_L, & (\overline{u_g})_L, & (u_b)_L, \\ (d_r)_L, & (u_g)_L, & (\overline{u_b})_L. \end{array} \tag{18}$$

The roots pointing in the direction of the weight $\omega_2$ can be determined as

$$\begin{array}{lll} \alpha_1 + \alpha_2 + \alpha_3, & \alpha_2, & \alpha_1 + \alpha_2, \\ \alpha_2 + \alpha_3, & \alpha_1 + \alpha_2 + \alpha_3 + \alpha_4, & \alpha_2 + \alpha_3 + \alpha_4 \end{array} \tag{19}$$

so the vectors in the direction to $(u_r)_L$ from the remaining particles in (18) are the gauge bosons corresponding to the roots in (19)

$$\begin{array}{lll} Y_r, & \overline{X_b}, & G_{r\bar{b}}, \\ W^+, & G_{r\bar{g}}, & \overline{X_g}\ . \end{array} \tag{20}$$

The other possibilities are depicted in Fig. 8.

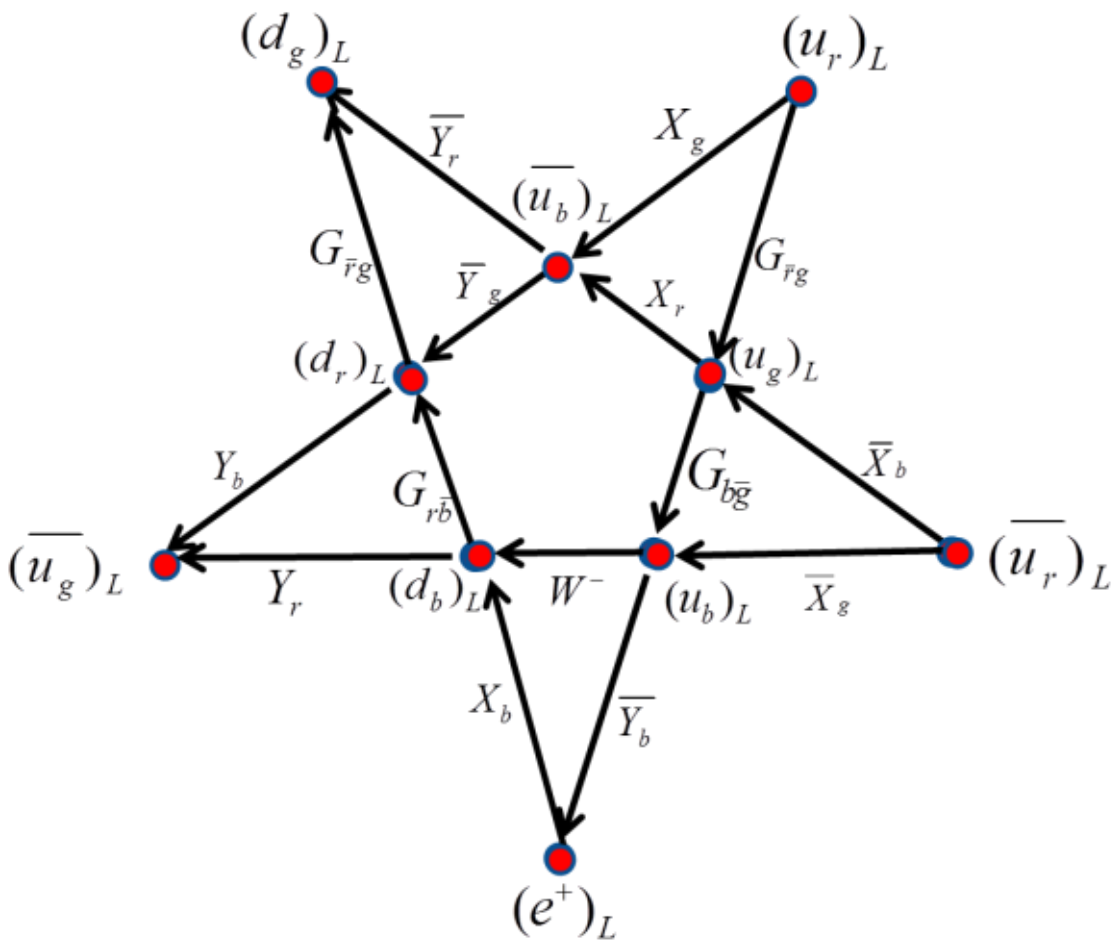

Fig. 8. First generation of the quark-lepton families in the representation $\underline{10}$ and the gauge bosons.

Let us have a closer look at one of the Delone cells. Consider the 5-cell represented by the root vectors including the origin $\alpha_1, \alpha_1+\alpha_2, \alpha_1+\alpha_2+\alpha_3, \alpha_1+\alpha_2+\alpha_3+\alpha_4, 0$. The first four roots correspond to the gauge bosons in the order $X_r$, $G_{r\bar{b}}, Y_r, G_{r\bar{g}}$. The 0 vector can be taken as any one of the four gauge bosons $\frac{1}{\sqrt{2}}(G_{r\bar{r}}-G_{b\bar{b}}), \frac{1}{\sqrt{6}}(G_{r\bar{r}}+G_{b\bar{b}}-G_{g\bar{g}}), W_3$ and $B$. In this 5-cell let us consider the edges from $\alpha_1$ to the rest. They are respectively $-\alpha_2$, $-(\alpha_2+\alpha_3), -(\alpha_2+\alpha_3+\alpha_4), \alpha_1$ which correspond to the gauge bosons $X_b, W^-, X_g, X_r$ .The diagrammatic representation of the gauge bosons and their interactions in the Delone 5-cell is depicted in Fig. 9.

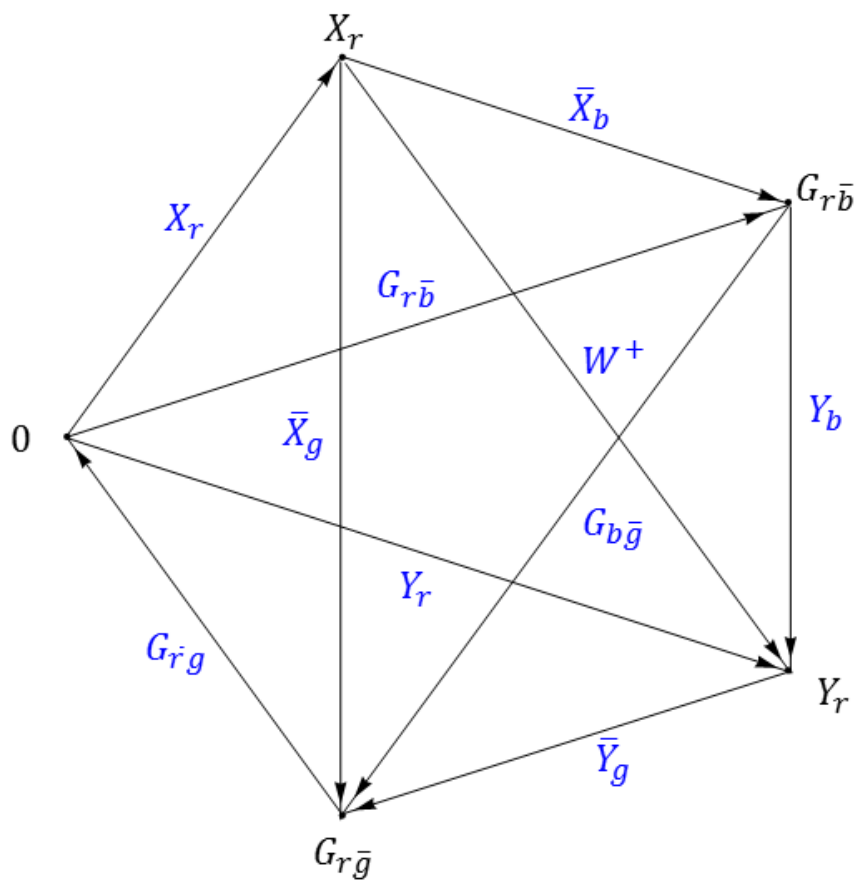

Fig. 9. Diagrammatic representation of the gauge bosons interaction in Delone 5-cell.

We also display the gauge boson interactions by projection of the root polytope as shown in Fig. 10.

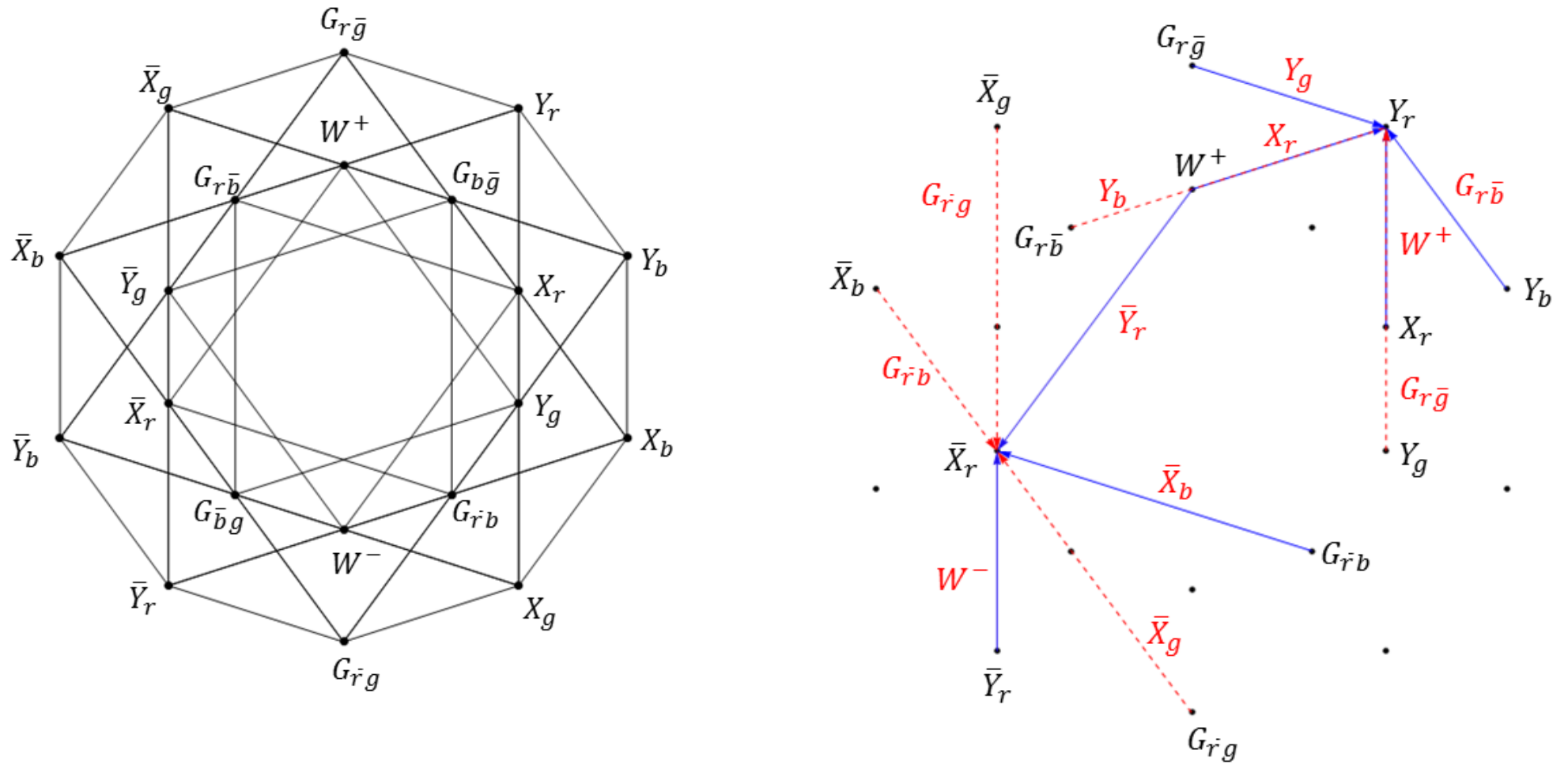

Fig.10. Diagrammatic representation of the gauge boson interactions in the root polytope. (a) Gauge bosons corresponding to the vertices, (b) Symbolic representations of two gauge bosons interacting with the nearest 6 gauge bosons.

## 4. Projections Of SU(5) Polytopes And Penrose-Like Tiling

Roger Penrose proposed an aperiodic tiling of the plane respecting 5-fold symmetry constructed with two unit cells consisting of thick and thin rhombs [23]. As we have seen above, the root lattice can be tiled by alternating Delone 5-cells and Delone rectified 5-cells. The 5-cell consists of tetrahedra whose faces are equilateral triangles. Similarly, the rectified 5-cell is made of tetrahedra and octahedra whose faces are all made of equilateral triangles. Projections of the 2D faces of the Delone cells onto the Coxeter plane lead to two types of isosceles triangles called Robinson triangles. If the equal edges are of unit length 1 then the third edge is either the golden ratio $\tau$ or its inverse $-\sigma$. The projection would lead to a 5-fold symmetric aperiodic tiling by Robinson triangles. For a different discussion on the $A_4$ projection see for instance Kramer [24]. A similar projection has been studied earlier by Baake and his collaborators [25].

We will not discuss the tiling by the Robinson triangles but rather aperiodic tiling by the Voronoi cell. Recall that in Section 2 we have already noted that the facets of the Voronoi cell are rhombohedra whose 2D faces are rhombs with edge length $\frac{2}{\sqrt{5}}$ and the interior angles are about $75.5^{\circ}$ and $104.5^{\circ}$. By using the projected components of the weight vectors in (15) we can prove that the six rhombs project into 4 thick rhombs with interior angles $72^{\circ}$ and $108^{\circ}$ and 2 thin rhombs with interior angles $36^{\circ}$ and $144^{\circ}$ as shown in Fig. 11.

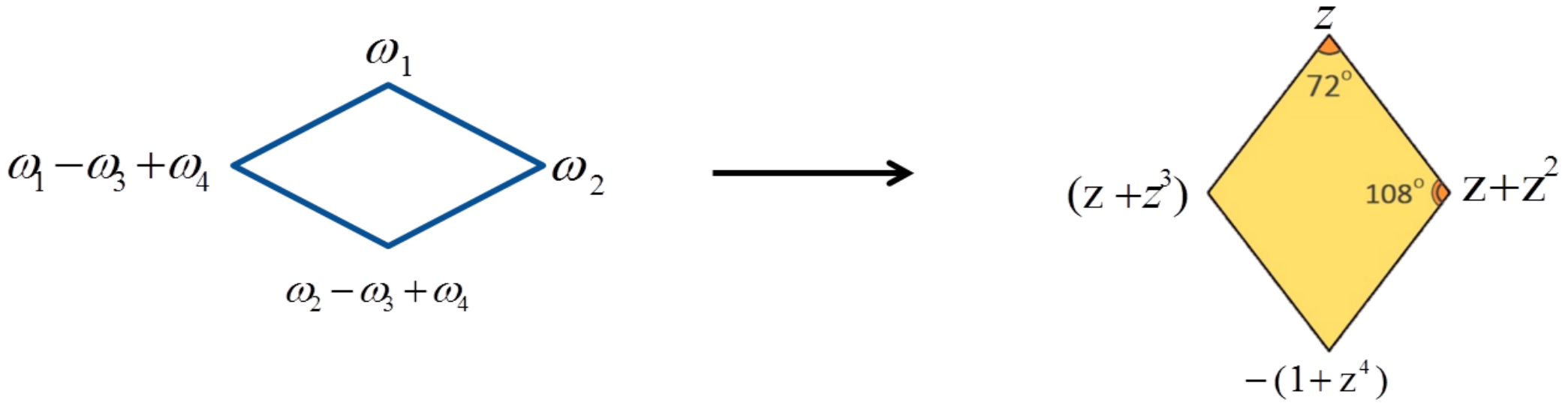

(a)

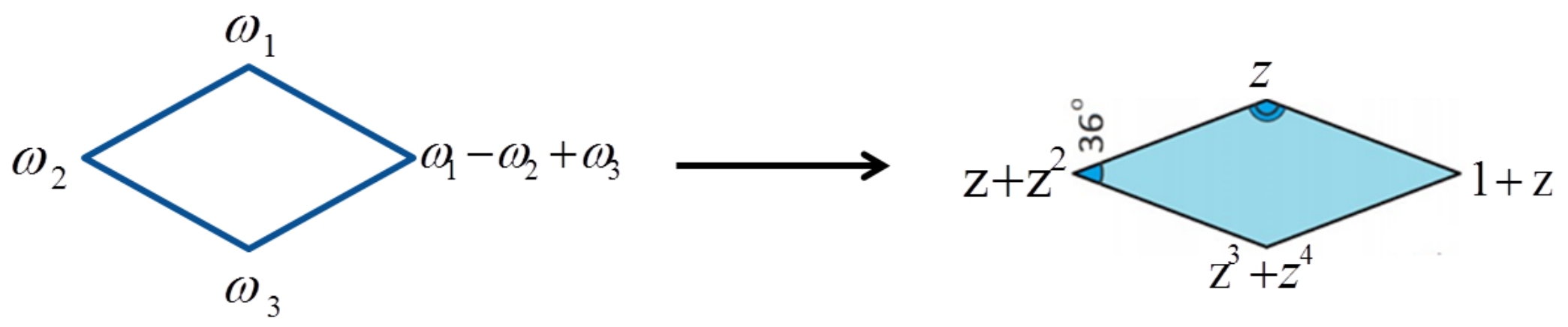

(b)

Fig. 11. Thick (a) and thin (b) rhombs from projection of rhombs of rhombohedral facets.

The vertices of the Voronoi cell decompose into three concentric decagons. As we have mentioned above the orbits $\underline{5}^*$ and $\underline{5}$ project on a circle with radius $\sqrt{\frac{2}{5}}$ as alternating points from two 5-cells. Similarly, half of the points of each orbits $\underline{10}$ and $\underline{10}^*$ project onto two different circles with radii $\sqrt{\frac{2}{5}}\,\tau$ and $-\sqrt{\frac{2}{5}}\,\sigma$ with alternating points from two orbits as shown in the Voronoi cell projected onto the following concentric decagons in Fig. 12. The orbits belonging to the vectors $\omega_1, \omega_2, \omega_3$ and $\omega_4$ are numbered as 1, 2, 3 and 4 respectively. Note that the radii of the concentric circles increase in the order $\tau^n, n = 1,2, \ldots$ .

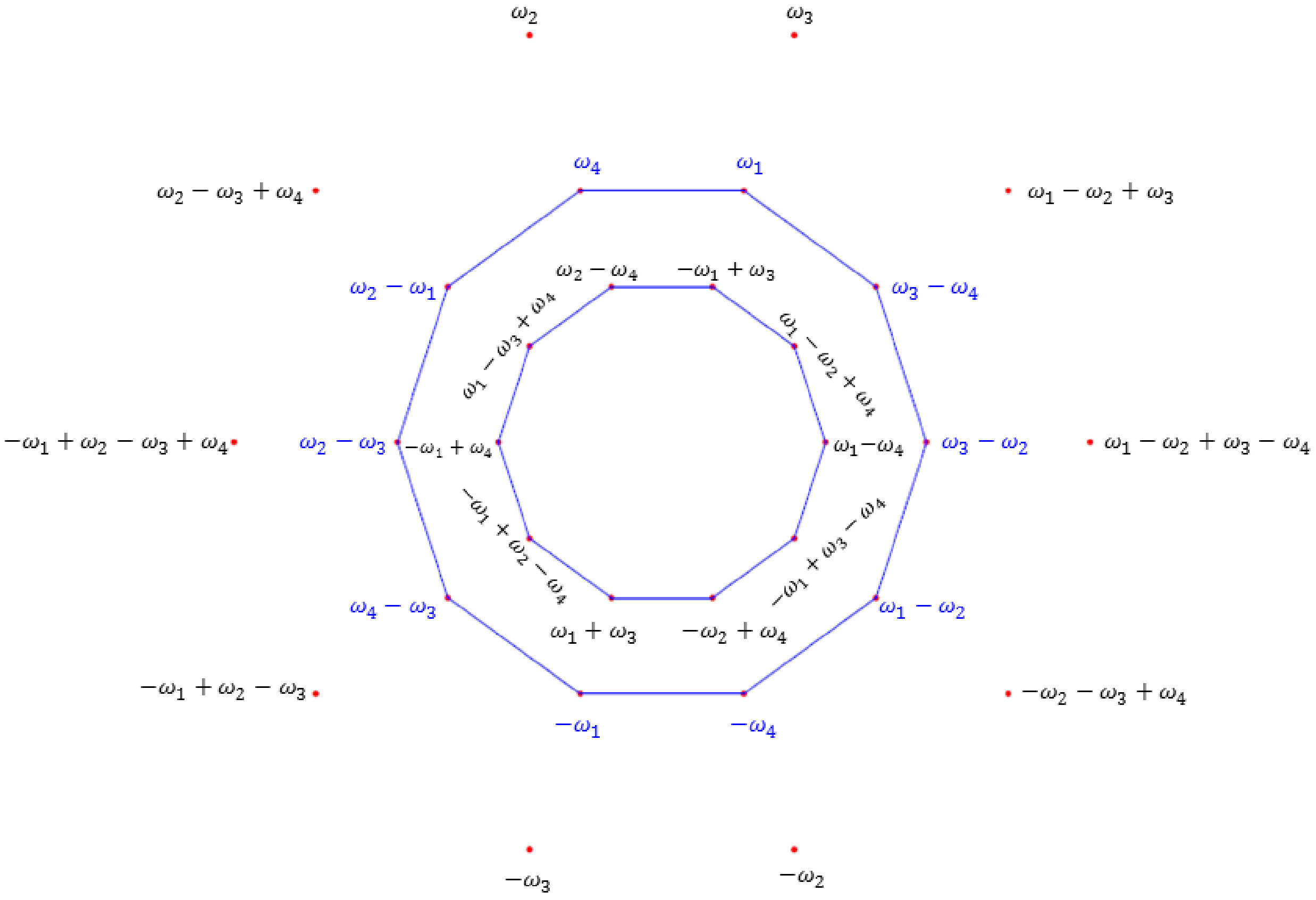

Fig. 12. Projection of the Voronoi cell.

In Figure 12 the edges of two inner decagons represent the projections of the 20 intersecting hyperplanes which go through the centers of the facets rhombohedra of the Voronoi cell $V(0)$. This implies that certain domains of the concentric circles overlap when Voronoi cell is projected. Our strategy is to tile the circular domain with thin and thick rhombs and then do reflections with respect to the allowed line segments. First we start to determine the tiling by fixing one point. Take for instance the point $\omega_1$ and determine the possible tiling around this fixed point. We have checked that the allowed tiles are those shown in Fig. 13.

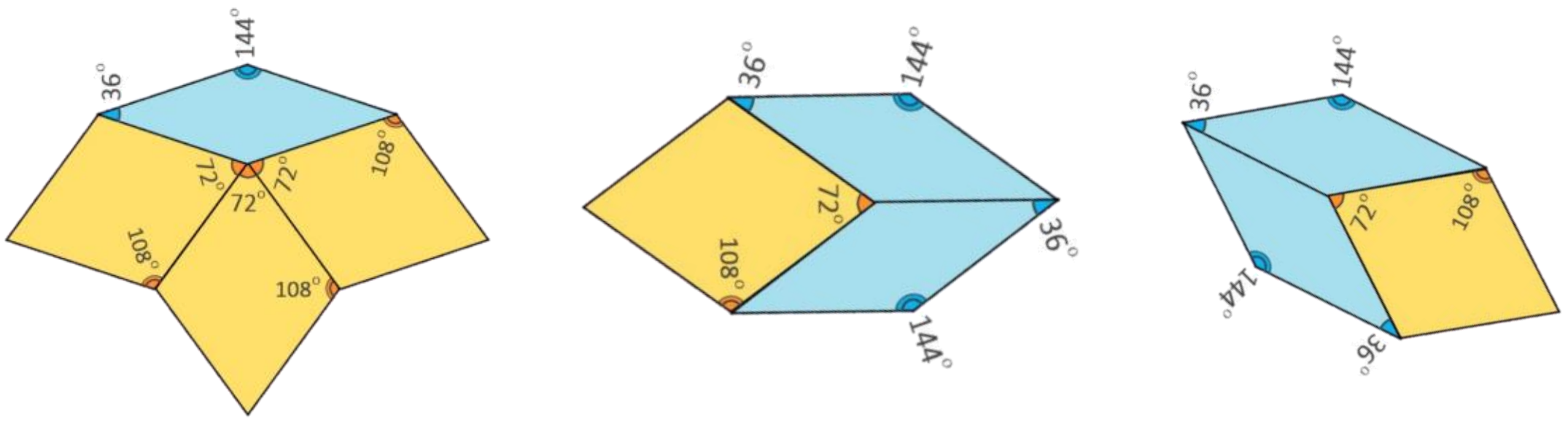

(a) (b) (c )

Fig. 13. Tiles centered at either $\omega_1$ or $\omega_4$ in the Voronoi cell.

By using the fixed point as $\omega_1$ or any other point in the same orbit with $\omega_1$ we can have four different tilings by projection of the Voronoi cell. They are illustrated in the Fig. 14 (a-d).

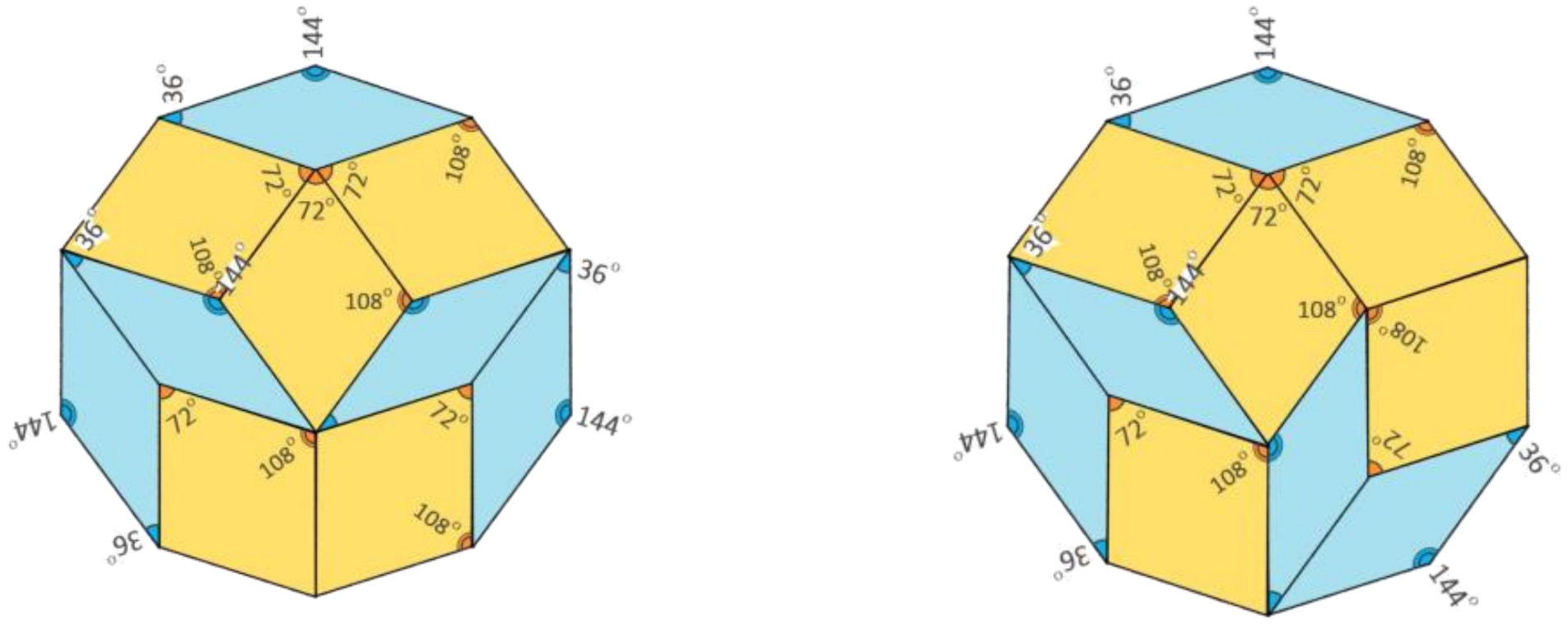

(a) (b)

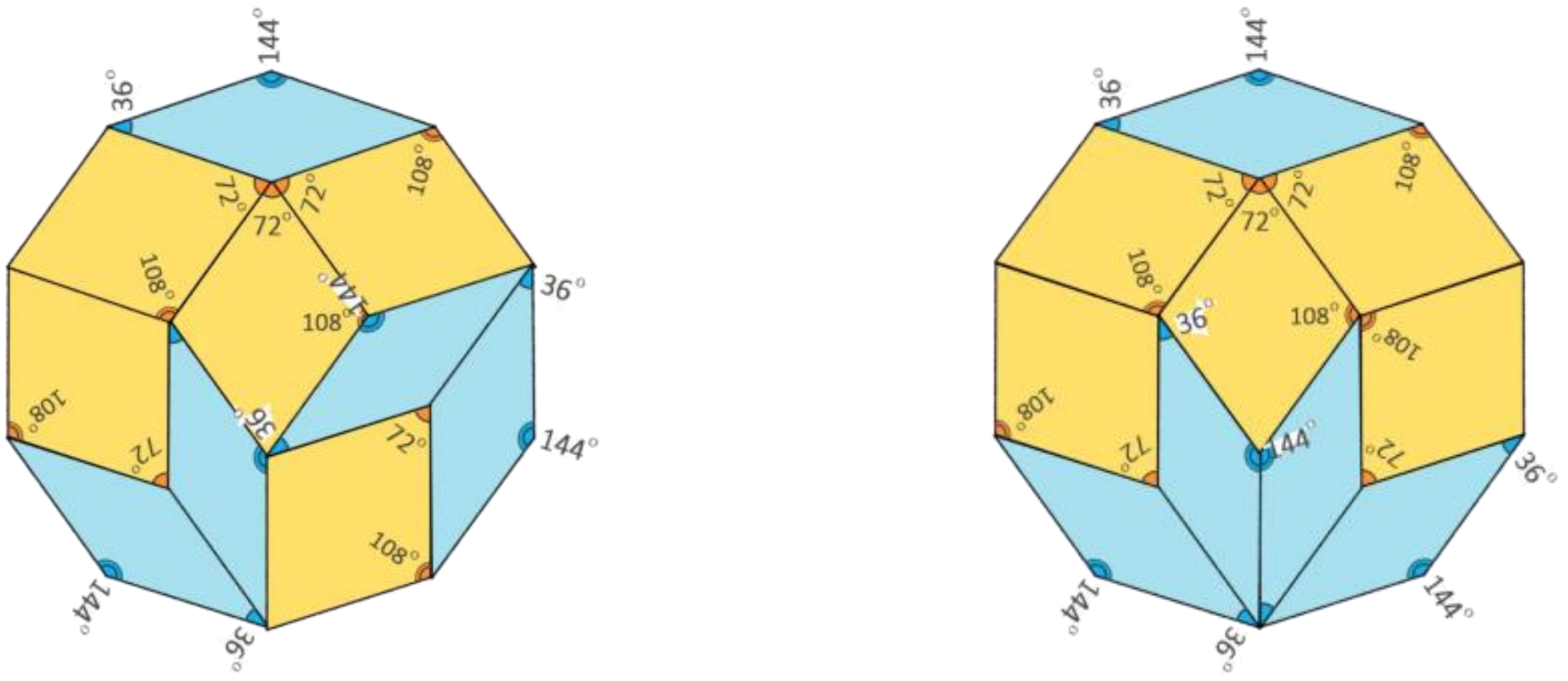

(c) (d)

Fig. 14. Possible projections of the Voronoi cell with thick and thin rhombs.

In all these tilings the number of thick and number of thin tiles are equal and therefore the ratio of the area covered by the thick rhombs to the area of the thin rhombs is the golden ration $\tau$. Some examples of tiling by affine reflections at the hyperplanes can be made with different tiles of Fig. 14. An example is shown in Fig. 15 which is simply made by folding along a line segment of the tile in Fig. 14(a). There are other techniques to obtain the tiles from higher dimensional lattices [31].

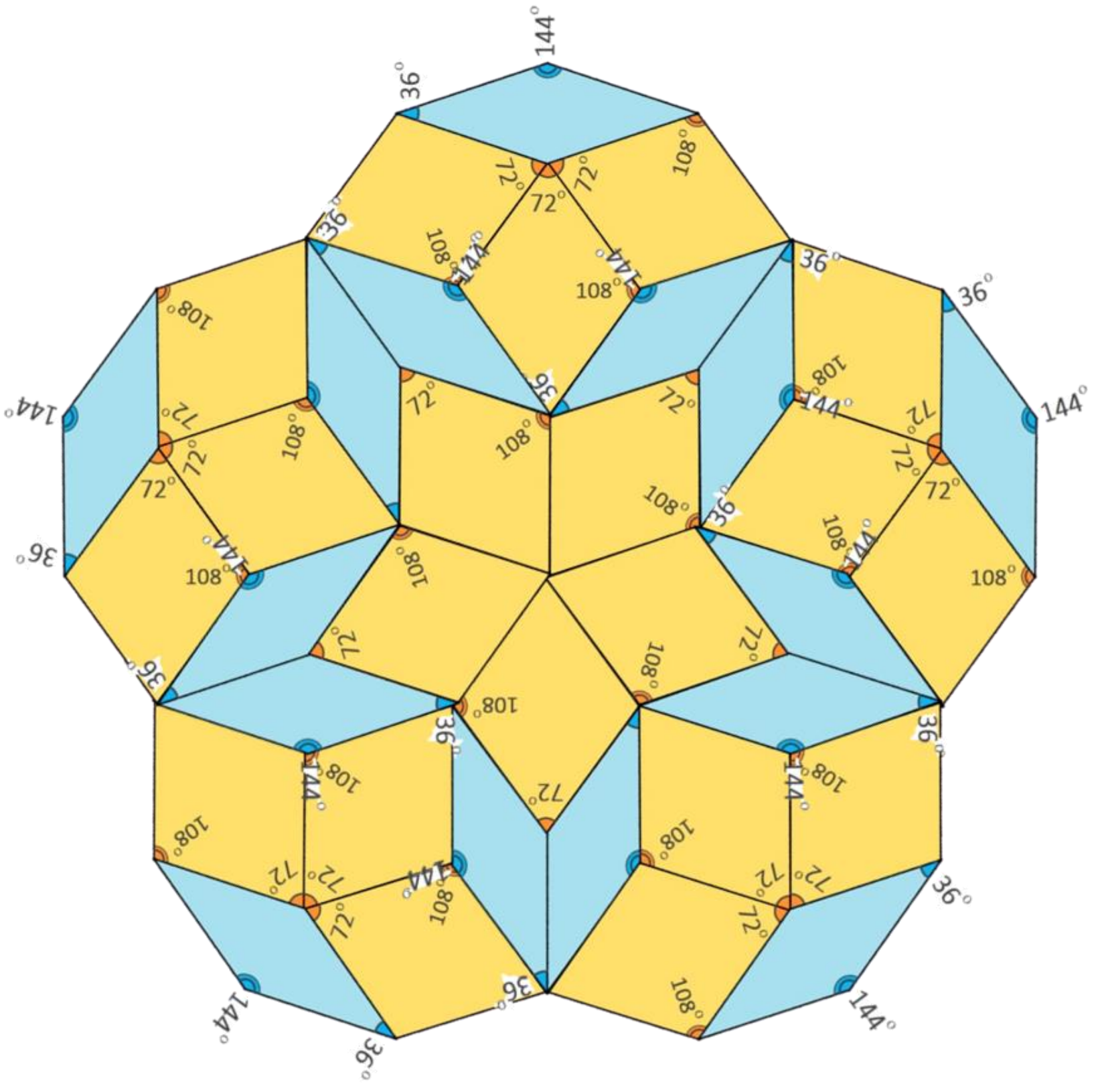

Fig. 15. Tiling with the projection of the Voronoi cell.

## 5. Concluding Remarks

We have presented a diagrammatic representation of the interaction of the $SU(5)$ particles and gauge bosons by the projected copies of the $SU(5)$ polytopes. We pointed out that projections of the congruent Voronoi cells correspond to the five-fold symmetric tiling of the Coxeter plane which is useful for the description of quasicrystallographic materials. This work can be extended to any GUT theory. An immediate concern could be the tilings obtained from the embedding $A_4 \subset D_5 \subset B_5$ which correspond to the gauge symmetry embedding $SU(5) \subset SO(10) \subset SO(11)$ where the relevant corresponding representations satisfy the relations $\underline{5}^* + \underline{10} + \underline{1} = \underline{16}$ and $\underline{16} + \underline{16}^* = \underline{32}$. The 32 weights of $B_5$ represent a cube in 5- dimensions and leads to 10-fold symmetric aperiodic tiling [21]. If one considers the projections of the $SO(10)$ polytopes into the Coxeter plane, we can prove that the aperiodic tiling displays the 8-fold symmetry. The 16 fermions project into two concentric octagons and the lines joining the points represent the gauge boson fermion interactions. In appendix A, we construct the quaternionic representation of the Coxeter-Weyl group $W(A_4) \subset W(H_4) \subset W(E_8)$ which is by itself very interesting to relate the $SU(5)$ lattice to a quasicrystallographic lattice $H_4$and then to the $E_8$ lattice. Appendix B lists the decompositions of the polytopes in terms of irreducible representations of the Coxeter-Weyl group $W(A_4)$.

## Appendix A: Quaternionic Representation Of The Coxeter-Weyl Group $W(A_4) \subset W(H_4)$

Let $q = q_o 1 + q_1 e_1 + q_2 e_2 + q_3 e_3$ be the real quaternion where the quaternionic imaginary units satisfy the relations

$$e_i e_j = -\delta_{ij} + \varepsilon_{ijk} e_k, i, j, k = 1,2,3, \quad \text{(A1)}$$

where $\delta_{ij}$ is the Kronecker symbol, meaning that it is equal to 1 if the indices are equal otherwise zero and $\varepsilon_{ijk}$ is the completely antisymmetric Levi-Civita symbol under the exchange of any two indices. It is well known that the root system of the quasicrystallographic Coxeter group $W(H_4)$ is represented by the 120 icosians $I = T + S$ [19] where two sets are given by

$$\begin{aligned} T &= \{ \pm 1, \pm e_1, \pm e_2, \pm e_3, \tfrac{1}{2}(\pm 1 \pm e_1 \pm e_2 \pm e_3)\}, \\ S &= \{\tfrac{1}{2}(\pm\tau \pm e_1 \pm \sigma e_3), \tfrac{1}{2}(\pm\tau \pm e_2 \pm \sigma e_1), \tfrac{1}{2}(\pm\tau \pm e_3 \pm \sigma e_2), \\ &\quad \tfrac{1}{2}(\pm\sigma \pm e_1 \pm \tau e_2), \tfrac{1}{2}(\pm\sigma \pm e_2 \pm \tau e_3), \tfrac{1}{2}(\pm\sigma \pm e_3 \pm \tau e_1), \\ &\quad \tfrac{1}{2}(\pm 1 \pm \tau e_1 \pm \sigma e_2), \tfrac{1}{2}(\pm 1 \pm \tau e_2 \pm \sigma e_3), \tfrac{1}{2}(\pm 1 \pm \tau e_3 \pm \sigma e_1), \\ &\quad \tfrac{1}{2}(\pm\sigma e_1 \pm \tau e_2 \pm e_3), \tfrac{1}{2}(\pm\sigma e_2 \pm \tau e_3 \pm e_1), \tfrac{1}{2}(\pm\sigma e_3 \pm \tau e_1 \pm e_2)\}. \end{aligned} \quad \text{(A2)}$$

The set $I$ also represents the vertices of 600-cell [26] where $T$stands for the binary tetrahedral group and represents the vertices of the 24-cell whose symmetry is the $W(F_4)$ of order 1152. The set of vertices $S$ represents the vertices of the snub 24-cell [27, 28]. Another representation of the icosians can be taken as $\tilde{I} = \tilde{T} + \tilde{S}$ where $\tilde{I}$ is obtained from $I$ by exchanging $\tau$ and $\sigma$. Consequently $T = \tilde{T}$ but $S \neq \tilde{S}$ . This is because binary icosahedral group has two 2-dimensional irreducible representations since the quaternions can be represented in terms of Pauli matrices $e_i = -i\sigma_i, i = 1, 2, 3$.

One possible representation of the simple roots of $SU(5)$ shown in Fig. 2 in terms of quaternions can be taken as

$$\alpha_1 = -\sqrt{2},\ \alpha_2 = \frac{1}{\sqrt{2}}(1 + e_1 + e_2 + e_3),\ \alpha_3 = -\sqrt{2}e_1,\ \alpha_4 = \frac{1}{\sqrt{2}}(e_1 - \sigma e_2 - \tau e_3). \quad \text{(A3)}$$

The scaled first three simple roots belong to the set $T$ and the scaled fourth simple root is an element of the set $S$.

The reflection generators $r_i = \frac{1}{2}[\alpha_i, -\alpha_i]^*,\ (i = 1, 2, 3, 4)$ generate the Coxeter-Weyl group $W(A_4)$. The bracket in the reflection generator means

$$r_i \mu = -\frac{\alpha_i}{\sqrt{2}} \bar{\mu} \frac{\alpha_i}{\sqrt{2}} := \frac{1}{2}[\alpha_i, -\alpha_i]^* \mu \quad \text{(A4)}$$

where $\mu$ is an arbitrary quaternion and $\bar{\mu}$ its quaternion conjugate. The group $W(A_4) \approx S_5$ with the simple roots above can be represented as

$$W(A_4) \approx S_5 = [p, \bar{c}\bar{\tilde{p}}c] + [p, c\bar{\tilde{p}}c]^*, p \in I, c = \frac{1}{\sqrt{2}}(e_2 - e_3). \quad \text{(A5)}$$

The set of elements $[p, \bar{c}\bar{\tilde{p}}c]$ is isomorphic to the group of even permutations of five objects (five vertices of the 5-cell) and represents the proper rotation subgroup of $W(A_4)$
where $\sim$ is defined as above.

The five-fold rotation operator ( the Coxeter element) in (12) can be written in the quaternionic basis as

$$R = [t, \bar{c}\bar{\tilde{t}}c], \text{ where } \quad t = \frac{1}{2}(-\sigma + e_2 + \tau e_3). \qquad \text{(A6)}$$

The extension of the group $W(A_4)$ by the Dynkin diagram symmetry is the automorphism group which can be compactly written as

$$Aut(W(A_4)) \approx S_5 := C_2,\ [p, \pm\bar{c}\bar{\tilde{p}}c] + [p, \pm c\bar{\tilde{p}}c]^*, \qquad \text{(A7)}$$

a group of order 240.

We have shown [29] that the group $W(A_4)$ is a maximal subgroup in the Coxeter group $W(H_4)$ with index 120. In other words, the group $W(A_4)$ can be embedded in the group $W(H_4)$ 120 different ways. In the notation of (A4) the group $W(H_4)$ can be represented as

$$W(H_4) = [p, q] + [p, q]^* \ , p, q \in I. \qquad \text{(A8)}$$

It is clear that the order of the group is $120 \times 120 = 14{,}400$. The group $W(H_4)$ is a maximal subgroup in $W(E_8)$ [30] the fact that can be used to project the $E_8$ lattice into quasicrystallographic $H_4$ lattice.

The representation of $W(A_4)$ in (A5) is just one choice out of 120 possibilities. For an arbitrary embedding one can write

$$W(A_4) = [p, \bar{\tilde{t}c}\bar{\tilde{p}}\tilde{t}c] + [p, tc\bar{\tilde{p}}\tilde{t}c]^*, p, t \in I. \qquad \text{(A9)}$$

Here 120 quaternions $tc \in I' = T' + S'$ where $I'$ represents the 600-cell in a different basis with $T' = Tc, S' = Sc$ representing a 24-cell ($T'$) and the snub 24-cell ($S'$) respectively.

## Appendix B: Decomposition Of SU(5) Polytopes In Terms Of Irreducible Representation Of $W(A_4)$

The character table of the group $W(A_4 \approx S_5$ has 7 irreducible representations of dimensions $1, 4, 5, 6, 5', 4', 1'$as shown in Table 5.

Table 5. The character table of the permutation group $S_5$.

| Classes | $(\mathbf{1^5})$ | $(\mathbf{1^3 2})$ | $(\mathbf{1^2 3})$ | $(\mathbf{12^2})$ | $(\mathbf{14})$ | (23) | (5) |
|---|---|---|---|---|---|---|---|
| #of elements/characters | 1 | 10 | 20 | 15 | 30 | 20 | 24 |
| χ(1) | 1 | 1 | 1 | 1 | 1 | 1 | 1 |
| χ(4) | 4 | 2 | 1 | 0 | 0 | -1 | -1 |
| χ(5) | 5 | 1 | -1 | 1 | -1 | 1 | 0 |
| χ(6) | 6 | 0 | 0 | -2 | 0 | 0 | 1 |
| χ($\mathbf{5'}$) | 5 | -1 | -1 | 1 | 1 | -1 | 0 |
| χ($\mathbf{4'}$) | 4 | -2 | 1 | 0 | 0 | 1 | -1 |
| χ($\mathbf{1'}$) | 1 | -1 | 1 | 1 | -1 | -1 | 1 |

The 5-cell, rectified 5-cell and the root polytope constitute the reducible representations of $S_5$ of dimensions 5,10 and 20 and they are decomposed in terms of the irreducible representations as

$$\underline{5} = \underline{1} + \underline{4}, \quad \underline{10} = \underline{1} + \underline{4} + \underline{5}, \quad \underline{20} = \underline{1} + 2(\underline{4}) + \underline{5} + \underline{6}.$$

## References

[1] S. Weinberg, A model of leptons, *Phys. Rev. Lett.* **19** (1967) 1264-1266.

[2] A. Salam, *Elementary Particle Theory: Relativistic Groups and Analyticity* (Proceedings Eighth Nobel Symposium, (1968) at Aspenasgerden, Lerum), Stockholm-New York: Almquist & Wiksell-Wiley Interscience, pp. 367-377.

[3] H. Fritzsch H, M. Gell-Mann and H. Leutwyler, Advantages of the color octet gluon picture, *Phys. Lett*. 47B (1973) 365-368.

[4] R. W. Carter, *Simple groups of Lie Type*, (John Wiley &Sons Ltd, 1972) pp. 364.

[5] H. Georgi H and S. L. Glashow, Unity of All Elementary-Particle Forces, *Phys. Rev. Lett.* **32** (1974) 438-441.

[6] H. Fritzsch and P. Minkowski, Unified interactions of leptons and hadrons, *Ann. Phys.* **93** (1975) 193–266.

[7] F. Gürsey, P. Ramond and P. Sikivie, A universal gauge theory model based on E6, *Physics Letters* B **60** (1976) 177-180.

[8] M. Koca, N. O. Koca and R. Koc, Group Theoretical Analysis of Quasicrystallography from Projections of Higher Dimensional Lattices $B_n$, *Acta Crystallographica* **A71** (2015) 175-185.

[9] N.O. Koca, M. Koca and R. Koc, 12-fold Symmetric Quasicrystallography from the lattices $F_4$, $B_6$, and $E_6$, *Acta Crystallographica* **A70** (2014) 605-615.

[10] M. Koca, N. O. Koca and M. Al-Ajmi, 4D-Polytopes and Their Dual Polytopes of the Coxeter Group W($A_4$) Represented by Quaternions, *Int. J. Geom. Methods Mod. Phys*. **9** (2012) 4.

[11] N. Arkani-Hamed and J. Trnka, *The Amplituhedron*, JHEP **10**, 030 (2014), arXiv:1312.2007(hep-th).

[12] M. Koca, R. Koc and M. Al-Ajmi, Quaternionic representation of the Coxeter group W(H4) and the polyhedra, *J. Phys. A: Math. Gen.* **39** (2006) 14047–14054.

[13] M. Koca, N. O. Koca and M. Al-Ajmi, Branching of the W ($H_4$) polytopes and their dual polytopes under the Coxeter groups W ($A_4$) and W($H_3$) represented by quaternions, *Tr. J. Phys*. **36** (2012) 309-333, arXiv: 1106.2957v1.

[14] H. S. M. Coxeter, The product of the generators of a finite group generated by reflections, *Duke Math J* **18** (1951) 765-782.

[15] H. S. M. Coxeter and W. O. J. Moser, *Generators and Relations for Discrete Groups* (Springer Verlag, 1965).

[16] N. Bourbaki, 1968 *Groupes et Algèbres de Lie. Chap. IV-VI* (Hermann, Paris. Russian translation 1972 Mir, Moscow; English translation, Springer, 2002).

[17] J. E. Humphreys, *Reflection Groups and Coxeter Groups* (Cambridge: Cambridge University Press, 1990) pp 204.

[18] R. Slansky, Phys. Rep. **C79**, 1 (1981); for an excellent technical review of the Lie Algebras see for instance H. Georgi, *Lie Algebra in Particle Physics* (2nd edition, Perseus Books, Reading Massachusetts, 1999).

[19] J. H. Conway and N. J. A. Sloane, *Sphere Packings, Lattices and Groups*, (3rd edition,Springer-Verlag, 1991) pp 703.

[20] M. Koca, R. Koc and M. Al-Ajmi, Polyhedra obtained from Coxeter groups and quaternions, *J. Math. Phys*. **48**, (2007) 113514-113527.

[21] N. G. de Brujin, Algebraic theory of Penrose's non-periodic tilings of the plane, *Proceedings of Koninklijke Nederlandse Akademie van Wetenschappen Series* A**84** (=*Indagationes Mathematicae,* **43**) (1981) 38-66.

[22] M. Koca, N. O. Koca and R. Koc, Affine $A_4$, Quaternions, and Decagonal Quasicrystals, *Int. J. Geom. Methods in Mod. Phys*. **11** (2014) 1450031.

[23] R. Penrose, The role of aesthetics in pure and applied mathematical research, *Bulletin of Inst. Math. and its Appl*. **10** (1974) 266-271.

[24] P. Kramer, Gateways towards quasicrystals, *J. Phys: Math. Gen.* **A33** (2000) 7885.

[25] M. Baake, P. M. Kramer, M. Scholtman and D. Zeidler, Planar patterns with fivefold symmetry as sections of periodic structures in 4–space, *Int. J. Mod. Phys.* **B4** (1990) 2217-2267.

[26] M. Koca, R. Koc and M. Al-Ajmi, Group theoretical analysis of 600-cell and 120-cell 4D polytopes with quaternions, *J. Phys. A: Math. And Theor*. **40** (2007) 7633–7642.

[27] M. Koca, M. Al-Ajmi and N. O. Koca, Quaternionic representation of snub 24-cell and its dual polytope derived from $E_8$ root system, *Linear Alg. Appl*. **434** (2011) 977-989.

[28] M. Koca, N. O. Koca and M. Al-Barwani, Snub 24-Cell Derived from the Coxeter-Weyl Group W ($D_4$), *Int. J. Geom. Methods Mod. Phys.* **9** (2012) 8.

[29] M. Koca, R. Koc, M. Al-Barwani and S. Al-Farsi, Maximal subgroups of the Coxeter group W(H4) and quaternions , *Linear Alg. Appl.* **412** (2006) 441-452.

[30] M. Koca, R. Koc and M. Al- Barwani, Noncrystallographic Coxeter group H4 in E8, *J. Phys. A: Math. Gen.A* **34** (2001) 11201–11213.

[31] M. Senechal, *Quasicrystals and Geometry* (Cambridge University Press, Cambridge, 1995).